\documentclass[preprint,5p]{elsarticle}

\usepackage{amsmath,amssymb}
\usepackage{braket}

\newcommand{\D}{\mathrm{d}}
\newcommand{\Tr}{\text{Tr}}
\newcommand{\e}{\mathrm{e}}
\renewcommand{\vec}[1]{\mathbf{#1}}

\begin{document}


\title{Fock-space projection operators for semi-inclusive final states}
\author[man]{Robert Dickinson}
\ead{robert.dickinson-2@manchester.ac.uk}
\author[man]{Jeff Forshaw}
\ead{jeff.forshaw@manchester.ac.uk}
\address[man]{Consortium for Fundamental Physics,
  School of Physics and Astronomy,
 University of Manchester,\\
 Manchester M13 9PL,
 United Kingdom.}
\author[not]{Peter Millington\corref{corr}}
\ead{p.millington@nottingham.ac.uk}
\address[not]{School of Physics and Astronomy, University of Nottingham,\\ Nottingham NG7 2RD, United Kingdom.\vspace{3em}}
\cortext[corr]{Corresponding author}


\begin{abstract}
We present explicit expressions for Fock-space projection operators that correspond to realistic final states in scattering experiments. Our operators automatically sum over unobserved quanta and account for non-emission into sub-regions of momentum space.  
\end{abstract}

\begin{keyword}
quantum field theory, projection operators
\end{keyword}

\maketitle


\section{Introduction}

When calculating matrix elements for scattering processes, it is necessary to sum over all final states that contribute to an observable, which often necessitates summing over unmeasured quanta. The classic example is the computation of the cross-section for $e^+e^-~\to~\text{hadrons}$, in which infra-red singularities cancel  between the virtual gluon corrections and corresponding zero-energy real gluon emissions (at the level of the squared matrix element) by the Kinoshita-Lee-Nauenberg theorem~\cite{Kinoshita:1962ur,Lee:1964is} (see also Refs.~\cite{Bloch:1937pw,Weinberg:1965nx}). Infra-red divergences can in fact be avoided at the amplitude level~(see e.g. Refs.~\cite{Chung:1965zza,Kulish:1970ut,Catani:1984dp,Forde:2003jt}), by absorbing  unobserved emissions into a re-definition of the asymptotic states. In this paper, we instead pursue the direct calculation of probabilities and focus on effect operators that correspond to the measurement of general semi-inclusive final states. These effect operators have the virtue that unobserved emissions simply do not enter the calculation.

The probability $\mathbb{P}$ that a system, described by some density operator $\rho$, will register an outcome, described by some effect operator $E$, is
\begin{equation}
	\mathbb{P} \ =\  \Tr (E \rho)~.
\end{equation}
Furthermore, if the measurement is performed at time $t_f$ and the system is known to be described at time $t_i$ by the density operator $\rho_i$ then, in the Interaction Picture, 
\begin{equation}
\rho_f \ =\  U{\!}_{fi}\, \rho_i\, U^\dag{\!}_{fi}~, 
\end{equation}
where
\begin{equation}
U_{fi} \ = \ \mathrm{T} \exp\bigg(\,\tfrac{1}{i}\!\int_{t_i}^{t_f} \! \D t\; H_{\mathrm{int}}(t) \bigg)
\end{equation}
is the unitary time-evolution operator and $H_\text{int}$ is the interaction Hamiltonian. If the initial state is a pure state, i.e.~$\rho_i=\ket{i}\!\bra{i}$, the probability takes the form
\begin{equation}
	\mathbb{P} \ =\  \bra{i} (U^\dag{\!}_{fi}\,E\,U{\!}_{fi}) \ket{i}\;.
\label{eq:prob}
\end{equation}
Of course, if the measurement also corresponds to a pure state, i.e.~$E=\ket{f}\!\bra{f}$, we obtain the usual squared matrix element
\begin{equation}
	\mathbb{P} \ =\  |\bra{f} U_{fi} \ket{i}|^2~.
\end{equation}
However, we may compute Eq.~(\ref{eq:prob}) directly by treating $E$ as an operator. We then view Eq.~\eqref{eq:prob} as an ``in-in'' expectation value, which can be written in the form \cite{Dickinson:2016oiy}
\begin{equation}
\label{eq:probfull}
\mathbb{P}\ =\ \sum_{j\,=\,0}\int_{t_i}^{t_f}{\rm d}t_1\,{\rm d}t_2\,\dots\,{\rm d}t_j\;\Theta_{12\dots j}\braket{i|\mathcal{F}_j|i}\;,
\end{equation}
where
\begin{subequations}
\begin{align}
\mathcal{F}_0\ &=\ E\;,\\
\mathcal{F}_j\ &=\ \tfrac{1}{i}\Big[\mathcal{F}_{j-1},H_{\rm int}(t_j)\Big]\;,
\end{align}
\end{subequations}
and $\Theta_{ijk\dots }\equiv 1$ if $t_i>t_j>t_k\dots$ and zero otherwise.

Whilst the explicit consideration of effect operators is ubiquitous in the description of measurement processes in quantum mechanics, they have, to our knowledge, been ignored in the context of particle physics. In what follows, we will present expressions for effect operators corresponding to general semi-inclusive measurements.

Our operators will be projection operators in Fock space and they all have the feature that unobserved quanta do not appear explicitly. For example, the effect operator corresponding to the inclusive cross-section for $e^+e^-\ \to$ one $q\,\bar{q}$ pair + anything is simply
\begin{equation}
E\ =\ \ket{q,\bar{q}}\!\bra{q,\bar{q}}\:\otimes\:\mathbb{I}_{e^+}\:\otimes\:\mathbb{I}_{e^-}\:\otimes\:\mathbb{I}_{\gamma}\:\otimes\:\mathbb{I}_g\;,
\end{equation}
where the sums over unobserved final-state electrons, posi\-trons, photons and gluons appear as unit operators in their respective Fock spaces, which trivially commute through the structure in Eq.~\eqref{eq:probfull}. These implicit summations over unobserved quanta are not present at the amplitude level, and this motivates further development of techniques along the lines of Ref.~\cite{Dickinson:2016oiy} aimed at directly computing probabilities in quantum field theory.

\section{Projection operators in Fock space:\\ bosonic case}

It is a well-known result in quantum optics that the vacuum projection operator can be written as the exponential of the photon number operator (see e.g.~Refs.~\cite{Louisell,Fan,Blasiak}):
\begin{align}
	E^{(0)}_{\mathbb{R}^3} \ &\equiv \ \mathbb{I} \;+\; \sum_{j\,=\,1}^\infty \frac{(-1)^j}{j!} : \big(N_{\mathbb{R}^3}\big)^j : \nonumber\\
&=\ : \e^{-N_{\mathbb{R}^3}} : \nonumber \\ &=\  \ket{0}\bra{0}~,
\label{eq:0psipo}
\end{align}  
where the number operator
\begin{equation}
\label{eq:NR}
N_{\mathcal{R}} \ \equiv\  \sum_\lambda \int_{\mathcal{R}} \frac{\D^3\mathbf{k}}{(2\pi)^3 2E}\; a_\lambda^\dag(\vec{k})a_\lambda(\mathbf{k})
\end{equation}
counts the number of quanta in a region $\mathcal{R}$ of momentum space, i.e.
\begin{equation}
N_{\mathcal{R}}\,\ket{\mathbf{k}_1\ldots\mathbf{k}_N} \ =\ n\,\ket{\mathbf{k}_1\ldots\mathbf{k}_N}~,
\end{equation}
where
\begin{equation}
n \ =\ \sum_{a\,=\,1}^N\mathbf{1}_{\mathcal{R}}(\mathbf{k}_a)
\end{equation}
and $\mathbf{1}_A(x)$ denotes the indicator function of set $A$, which is 1 if $x\in A$ and 0 otherwise. The colons indicate normal ordering. The sum is over all physical polarizations $\lambda$, if the projection is to be independent of polarization, or it could be over some subset of all allowed polarizations. Moreover, the region of momentum space need not be common to all polarizations, i.e.~$\mathcal{R}\to\mathcal{R}_{\lambda}$. For ease of notation, we suppress the polarization indices that are needed to fully specify Fock states.

Whilst $E^{(0)}_{\mathbb{R}^3}$ is the projection operator corresponding to zero quanta (anywhere in configuration space), we can show that 
\begin{equation}
E^{(0)}_{\mathcal{R}}\ \equiv\ : \e^{-N_{\mathcal{R}}} : 
\end{equation}
is the projection operator corresponding to zero quanta in the region $\mathcal{R}$, i.e.
\begin{align}
E^{(0)}_{\mathcal{R}}\,\ket{\mathbf{k}_1\ldots\mathbf{k}_N} 
&=\  \begin{cases}
\ket{\mathbf{k}_1\ldots\mathbf{k}_N} &\text{if zero quanta in } \mathcal{R}~, \\
0 & \text{otherwise~.}
\end{cases} \label{eq:sud0}
\end{align}
The proof of this result, and of those that follow, is contained in the appendix. For $\mathcal{R}=\mathbb{R}^3$, we project out the vacuum state, as in Eq.~\eqref{eq:0psipo}. For the opposite extreme $\mathcal{R}=\emptyset$ (the empty set), the effect operator is just the unit operator, i.e.~$E^{(0)}_{\emptyset}=\mathbb{I}$, and the measurement is inclusive over all final states [cf.~Eq.~\eqref{eq:prob}]. Everywhere in between, we automatically sum over states with zero quanta inside $\mathcal{R}$ and \emph{any number} of quanta outside $\mathcal{R}$.

These non-emission operators are specific cases of a more general projection operator:
\begin{align}
E^{\{j_a\}}_{\{\mathcal{R}_a\subseteq \mathcal{R}\}} \ &\equiv\ : \Bigg[\prod_a \frac{1}{j_a!}\big(N_{\mathcal{R}_a}\big)^{j_a} \Bigg]\, \e^{-N_{\mathcal{R}}} :~.
\label{eq:jpsipo}
\end{align}
This operator projects onto the subspace of states in which exactly $\sum_a j_a$ quanta have momenta in $\mathcal{R}$, distributed so that exactly $j_a$ quanta have momenta in each disjoint subset $\mathcal{R}_a \subseteq\mathcal{R}$. Again, there is no restriction on quanta lying outside of $\mathcal{R}$.
The special case of
\begin{align}
E^{(j)}_{\mathcal{R}} \ &\equiv\ : \frac{1}{j!}\big(N_{\mathcal{R}}\big)^{j}\, \e^{-N_{\mathcal{R}}} : 
\end{align}
projects onto exactly $j$ particles in $\mathcal{R}$ and resembles the operator form of the photon counting distribution in quantum optics (see e.g.~Ref.~\cite{Klauder}).

To illustrate Eq.~(\ref{eq:jpsipo}), we might consider the simple case where one quantum has momentum in the range $\mathbf{k} \to \mathbf{k}+\D^3 \mathbf{k}$ and there are no other quanta anywhere, i.e. $\mathcal{R} = \mathbb{R}^3$. In this case, the projection operator is
\begin{align}
E^{(1)}_{\mathcal{R}_1 \subset \mathbb{R}^3} \ &=\ : N_{\mathcal{R}_1} \e^{-N_{\mathbb{R}^3}}: \nonumber\\ & =\  \frac{\D^3 \vec{k}}{(2\pi)^3 2E} : a^\dag(\vec{k}) a(\vec{k}) \ket{0} \bra{0} : \nonumber\\ &= \ \frac{\D^3 \vec{k}}{(2\pi)^3 2E} \ket{\vec{k}} \bra{\vec{k}}~.
\end{align}
With $j_a=1$ $\forall$ $a$, Eq.~(\ref{eq:jpsipo}) could be employed in situations where the observable final state has the form of $n$ particles with given momenta $\mathbf{k}_a  \to \mathbf{k}_a+\D^3 \mathbf{k}_a$, accompanied by any number of undetectable particles below a given energy and/or transverse momentum threshold.

Since these projection operators share a common eigenbasis --- the Fock basis --- they mutually commute and can be combined straightforwardly. For example,
$E^{(j)}_{\mathcal{R}_1} E^{(k)}_{\mathcal{R}_2}$ projects onto states with exactly $j$ quanta in $\mathcal{R}_1$ \emph{and} exactly $k$ quanta in $\mathcal{R}_2$, regardless of whether $\mathcal{R}_1$ and $\mathcal{R}_2$ are disjoint.

We may now construct a projection operator $\D E_{\mathcal{R},v_n}(V)$ for an $n$-particle final state satisfying a constraint of the form $V \leq v_n(\mathbf{k}_1,\ldots,\mathbf{k}_n) \leq V+\D V$, which is symmetric under interchange of any two momenta, and inclusive of particles outside region $\mathcal{R}$. In the low-density regime in which Fock state occupation numbers rarely exceed unity, this is
\begin{align}
\frac{\D E_{\mathcal{R},v_n}}{\D V}\ &= \  \Bigg[\prod_{i\,=\,1}^n  \int_{\mathcal{R}} \frac{\D^3\mathbf{k}_i}{(2\pi)^3 2E_i} \Bigg] \frac{1}{n!}\,\delta(v_n(\{\mathbf{k}_i\})-V)\nonumber\\&\qquad\qquad \times\: :\Bigg[\prod_{i\,=\,1}^n a^\dag(\mathbf{k}_i)a(\mathbf{k}_i) \Bigg] \e^{-N_{\mathcal{R}}} :~.
\end{align}
Where a single choice of particle number $n$ is not appropriate, we may bring the particle number into the constraint function $v\big( \{\mathbf{k}_i\};n \big) = v_n(\mathbf{k}_1,\ldots,\mathbf{k}_n)\;\forall\;n$, and define
\begin{align}
\frac{\D E_{\mathcal{R},v}}{\D V} \ =\  \sum_n \frac{\D E_{\mathcal{R},v_n}}{\D V}~.
\end{align}


\section{Projection operators in Fock space:\\ fermionic case}

The projection operators for fermions are analogous to the bosonic case. We may regard the sum over $\lambda$ in Eq.~\eqref{eq:NR} to be inclusive of particle ($b_s^{\dag}(\vec{k})b_s(\vec{k})$), anti-particle ($d_s^{\dag}(\vec{k})d_s(\vec{k})$) and spin states (indexed by $s$), i.e.~$N_{\mathcal{R}} \to N_{\mathcal{R}}+\bar{N}_{\bar{\mathcal{R}}}$, where
\begin{subequations}
\begin{align}
N_{\mathcal{R}}\ =\  \sum_s \int_{\mathcal{R}} \frac{\D^3\mathbf{k}}{(2\pi)^3 2E}\;b_s^\dag(\vec{k})b_s(\mathbf{k})~,\\
\bar{N}_{\bar{\mathcal{R}}}\ =\ \sum_s \int_{\bar{\mathcal{R}}} \frac{\D^3\mathbf{k}}{(2\pi)^3 2E}\; d_s^\dag(\vec{k})d_s(\mathbf{k})~.
\end{align}
\end{subequations}
As was true of the polarization sum, the regions $\mathcal{R}$ and $\bar{\mathcal{R}}$ need not be common to all spin projections, i.e.~$\mathcal{R}\to\mathcal{R}_s$ and $\bar{\mathcal{R}}\to\bar{\mathcal{R}}_s$.
The anti-commutativity of the fermion creation and annihilation operators is accounted for in the definition of normal ordering:
\begin{subequations}
\begin{align}
:b_s^{\dag}(\mathbf{k})b_s(\mathbf{k}):\ =\ +\:b^{\dag}_s(\mathbf{k})b_s(\mathbf{k})~,\\
:b_s(\mathbf{k})b_s^{\dag}(\mathbf{k}):\ =\ -\:b^{\dag}_s(\mathbf{k})b_s(\mathbf{k})~,
\end{align}
\end{subequations}
with analogous expressions holding for the anti-fermion operators $d^{\dag}_s(\vec{k})$ and $d_s(\vec{k})$.

For a general product of $j$ operators, we find
\begin{align}
\label{eq:fermionprod}
&:\prod_{a\,=\,1}^{j}b^{\dag}_{s_a}(\mathbf{k}_a)b_{s_a}(\mathbf{k}_a):\nonumber\\&\qquad\qquad =\ (-1)^{j(j-1)/2} \prod_{a\,=\,1}^jb^{\dag}_{s_a}(\mathbf{k}_a)\prod_{b\,=\,1}^jb_{s_b}(\mathbf{k}_b)~.
\end{align}
The normal ordering has given rise to an overall factor of $(-1)^{j(j-1)/2}$. However, after acting on a given state with the annihilation operators, the order of the creation operators is reversed relative to the original state. Using anti-commutation to recover the original order, we pick up an additional factor of $(-1)^{j(j-1)/2}$, with the result that there is no overall sign relative to the bosonic case. We can account for this directly at the level of Eq.~\eqref{eq:fermionprod} by re-ordering the creation operators, picking up the same additional factor of $(-1)^{j(j-1)/2}$:
\begin{equation}
:\prod_{a\,=\,1}^{j}b^{\dag}_{s_a}(\mathbf{k}_a)b_{s_a}(\mathbf{k}_a):\ =\ \prod_{a\,=\,j}^1b^{\dag}_{s_a}(\mathbf{k}_a)\prod_{b\,=\,1}^jb_{s_b}(\mathbf{k}_b)~.
\end{equation}
The behaviour of the normal-ordered products of fermion number operators is therefore identical to that of the nor\-mal-ordered boson number operators described previously. This can also be understood by virtue of the fact that fermionic number operators are \emph{commutative} not anti-com\-mutative.

As an example, the operator projecting onto the subspace of states in which there are exactly $j$ fermions (of any spin) and zero anti-fermions in $\mathcal{R}$ is
\begin{align}
E^{(j,0)}_{\mathcal{R}}\ &\equiv\ :\frac{1}{j!}(N_{\mathcal{R}})^je^{-N_{\mathcal{R}}-\bar{N}_{\mathcal{R}}}:\nonumber\\& =\;:\frac{1}{j!}(N_{\mathcal{R}})^je^{-N_{\mathcal{R}}}:\otimes:e^{-\bar{N}_{\mathcal{R}}}:~.
\end{align}

In all cases, the projection operators of a given degree of freedom are built from the corresponding number operator. The results presented here may therefore be generalized readily to include additional gauge structure, multiple flavours or higher-spin representations, simply by accounting for summations over the additional quantum numbers. Since the number operators of different degrees of freedom mutually commute --- for fermions as well as bosons --- their projection operators may be combined straightforwardly by tensor multiplication. One can then imagine constructing semi-inclusive projection operators able to deal with final states of any content and complexity by combining those of different species across various disjoint and/or overlapping regions of momentum space.


\section{Conclusions}

These projection operators have the interesting property that unobserved quanta never appear in the calculation. This may have a significant impact upon the way in which we deal with infra-red divergences in gauge theories. In order to take advantage of this property, we must compute probabilities \emph{directly}, bypassing amplitude-level calculations altogether. Were we to revert to the latter, we would need to break the projection operators apart again, reintroducing the explicit sums over unobserved emissions that we intend to avoid. It remains to develop technology that makes tractable the explicit calculation of these probabilities, perhaps building on the results of Ref.~\cite{Dickinson:2016oiy} and the earlier ideas of Ref.~\cite{Dickinson:2013lsa} by exploiting the connection to the path-integral approach of the in-in (or closed-time-path) formalism.


\section*{Acknowledgements}

The work of PM is supported by STFC Grant No.\\ ST/L000393/1 and a Leverhulme Trust Research Leadership Award.


\appendix
\section{Proofs of results quoted in the main text}

It is useful to be able to compute the eigenvalues of normal-ordered products of number operators. The eigenvalue equations themselves have identical forms for bosonic and fermionic number operators, and we will suppress all but the momentum dependence of states for conciseness. The first non-trivial example is
\begin{equation}
\label{eq:N2}
: N_{\mathcal{R}_1}N_{\mathcal{R}_2} : \ket{\mathbf{k}_1\ldots\mathbf{k}_N} =\;(n_1 n_2-n_{12})\,\ket{\mathbf{k}_1\ldots\mathbf{k}_N}\,,\\
\end{equation}
where $n_i$ counts the number of quanta lying in $\mathcal{R}_i$ and $n_{12}$ counts the number of quanta lying in the overlapping region $\mathcal{R}_1 \cap \mathcal{R}_2$. Similarly,
\begin{align}
&: N_{\mathcal{R}_1}N_{\mathcal{R}_2}N_{\mathcal{R}_3} : \ket{\mathbf{k}_1\ldots\mathbf{k}_N} \;=\;(n_1 n_2 n_3- n_{12} n_3\nonumber\\&\qquad -n_{13} n_2-n_{23} n_1+2 n_{123})\,\ket{\mathbf{k}_1\ldots\mathbf{k}_N}\,.
\end{align} 
These are the simplest examples of the more general formula: 
\begin{align}
& : N_{\mathcal{R}_1}N_{\mathcal{R}_2} \ldots N_{\mathcal{R}_p} : \ket{\mathbf{k}_1\ldots\mathbf{k}_N} \nonumber\\&\qquad= 
\Bigg{[}\prod_{r\,=\,1}^{p} n_r \Bigg{]} \Bigg{[} 1 \nonumber\\&\qquad \qquad +\: \;\sum_{i < j} \frac{(-1)n_{ij}}{n_in_j} \nonumber\displaybreak \\ &\qquad\qquad  + \: \!\sum_{i < j < k} \frac{(-1)(-2)n_{ijk}}{n_in_jn_k} \nonumber\\ & \qquad\qquad +\: \sum_{i < j < k<l} \!\!\!\frac{(-1)(-2)(-3)n_{ijkl}}{n_in_jn_kn_l} +\ldots \nonumber \\ &
\qquad\qquad +\: \!\!\!\!\!\sum_{i < j, i<k< l} \!\!\!\frac{(-1)n_{ij}(-1)n_{kl}}{n_in_jn_kn_l} \nonumber\\ & \qquad\qquad +\: \!\!\!\!\sum_{i < j, k < l<m} \!\!\!\frac{(-1)(-2)n_{ijk}(-1)n_{lm}}{n_in_jn_kn_ln_m} \:+\: \ldots \nonumber \\ &\qquad\qquad  
+\: \frac{(-1)^{p-1} (p-1)! n_{12\ldots p}}{n_1n_2\ldots n_p}
\Bigg]
\ket{\mathbf{k}_1\ldots\mathbf{k}_N}\,,
\label{eq:partitionseval}
\end{align}
in which a sum is listed for every integer partition of $p$.

The eigenvalue of this normal-ordered product of number operators counts the total number of ways to select $p$ quanta from the set specified by the state $\ket{\mathbf{k}_1\ldots\mathbf{k}_N}$ such that one quantum is in each of the regions $\mathcal{R}_i$.
If the regions are nested, such that $\mathcal{R}_1 \subseteq \mathcal{R}_2 \subseteq \ldots$, Eq.~\eqref{eq:partitionseval} reduces to
\begin{equation}
:\prod_{i\,=\,1}^p (N_{\mathcal{R}_i}) : \ket{\mathbf{k}_1\ldots\mathbf{k}_N} = \Bigg[\prod_{i\,=\,1}^p \big(n_i-(i\!-\!1)\big)\Bigg]\ket{\mathbf{k}_1\ldots\mathbf{k}_N}\,,
\label{eq:nestedns}
\end{equation}
and, if all the $\mathcal{R}_i$ are identical, this becomes
\begin{equation}
:(N_{\mathcal{R}})^p : \ket{\mathbf{k}_1\ldots\mathbf{k}_N} = \begin{cases} \dfrac{n!}{(n-p)!} \,\ket{\mathbf{k}_1\ldots\mathbf{k}_N}&\text{if}\ n\geq p~, \\
0 &\text{otherwise~.} \end{cases}
\label{eq:identicalns}
\end{equation}

We now consider more than one sequence of nested regions in the case that the regions in different sequences are disjoint. If we have $j_1$ copies of region $\mathcal{R}_1$, $j_2$ copies of region $\mathcal{R}_2$ and so on, with $\mathcal{R}_i \cap \mathcal{R}_j = \emptyset\;\;\forall\; i\neq j$, then
\begin{align}
&:\prod_{a} (N_{\mathcal{R}_a})^{j_a}: \ket{\mathbf{k}_1\ldots\mathbf{k}_N} \nonumber\\ &\qquad  =\  \begin{cases} \displaystyle \Bigg[\prod_{a} \frac{n_a!}{(n_a-j_a)!}\Bigg]\ket{\mathbf{k}_1\ldots\mathbf{k}_N}&\text{ if }n_a\geq j_a \ \forall\ a~, \\
0 &\text{ otherwise~.} \end{cases}
\label{eq:bunchedns}
\end{align}
The product form of the eigenvalues is a consequence of the fact that the operator factorizes into mutually commuting operators of the form given in Eq.~(\ref{eq:identicalns}). 

Now we consider a set of disjoint regions within a superset. Let us augment the case of the previous paragraph with $k$ copies of a region $\mathcal{R} \supset \mathcal{R}_i$. After selecting $j_a$ particles from each disjoint region $\mathcal{R}_a$, the number of particles remaining in $\mathcal{R}$ is $n_x \equiv n-\sum_a j_a$. The number of ways of selecting these $k$ particles is then $n_x!/(n_x-k)!$, and
\begin{align}
&:\Bigg[\prod_{a} (N_{\mathcal{R}_a})^{j_a}\Bigg](N_{\mathcal{R}})^{k}: \ket{\mathbf{k}_1\ldots\mathbf{k}_N} \nonumber\\& = \begin{cases} \displaystyle \frac{n_x!}{(n_x - k)!}\Bigg[\prod_{a} \frac{n_a!}{(n_a-j_a)!}\Bigg]\!\ket{\mathbf{k}_1\ldots\mathbf{k}_N}&\begin{array}{l}\hspace{-0.9em}\text{if}\ n_a\geq j_a \ \forall\  a \\\hspace{-0.9em}\text{and}\ n_x\geq k~, \end{array} \\
0 &\ \ \hspace{-1em}\text{otherwise~.} \end{cases}
\label{eq:bunchedsupns}
\end{align}
These results for the action of normal-ordered products of the number operator are the key to proving the results quoted in the main text.

Specifically, using Eq.~(\ref{eq:identicalns}), we can go ahead and prove Eq.~(\ref{eq:sud0}): 
\begin{align}
E_{\mathcal{R}}^{(0)}\,\ket{\mathbf{k}_1\ldots\mathbf{k}_N} \ &=\
 \sum_{p\,=\,0}^{n} \frac{(-1)^p \, n!}{p!(n-p)!} \ket{\mathbf{k}_1\ldots\mathbf{k}_N}\nonumber\\
&\ =\  \lim_{x\,\rightarrow\, -1}(1+x)^{n}\,  \ket{\mathbf{k}_1\ldots\mathbf{k}_N}  \nonumber\\
&\ =\  \begin{cases}
\ket{\mathbf{k}_1\ldots\mathbf{k}_N} &\text{if } n=0~, \\
0 & \text{otherwise~.}
\end{cases}
\end{align}
A proof of Eq.~(\ref{eq:jpsipo}), using Eq.~(\ref{eq:bunchedsupns}) with $n_x \equiv n-\sum_a j_a$, runs as follows:
\begin{align}
&E^{\{j_a\}}_{\{\mathcal{R}_a\subseteq \mathcal{R}\}}\,\ket{\mathbf{k}_1\ldots\mathbf{k}_N} \nonumber\\ &=\ \sum_{k\,=\,0}^\infty \frac{(-1)^k}{k!} : \Bigg[\prod_a \frac{1}{j_a!}(N_{\mathcal{R}_a})^{j_a}\Bigg] (N_{\mathcal{R}})^k : \ket{\mathbf{k}_1\ldots\mathbf{k}_N} \nonumber\\
&=\  \displaystyle\sum_{k\,=\,0}^{n_x} \frac{(-1)^k}{k!} \frac{n_x!}{(n_x-k)!}\Bigg[\prod_a \frac{n_a!}{j_a!(n_a-j_a)!}\Bigg] \ket{\mathbf{k}_1\ldots\mathbf{k}_N} 
\end{align}
provided $n_a\geq j_a \;\forall a$, and zero otherwise. The eigenvalue may be written
\begin{equation}
\Bigg[\prod_a {n_a\choose j_a}\Bigg]\!\sum_{k\,=\,0}^{n_x} \!{n_x \choose k}(-1)^k  =  \Bigg[\prod_a {n_a\choose j_a}\Bigg] \!\lim_{x\,\rightarrow\, -1} (1+x)^{n_x} ~,
\end{equation}
which vanishes unless $n_x=0$. Since $n \geq \sum_a n_a \geq \sum_a j_a = n-n_x$, this implies $n_a=j_a \;\forall \;a$. Hence,
\begin{align}
&E^{\{j_a\}}_{\{\mathcal{R}_a\subseteq \mathcal{R}\}}\,\ket{\mathbf{k}_1\ldots\mathbf{k}_N} \nonumber\\&\ =\  \begin{cases}
\ket{\mathbf{k}_1\ldots\mathbf{k}_N} &\text{if } n_a = j_a \;\forall\; a \text{ and }n = \sum_a j_a~, \\
0 & \text{otherwise~.}
\end{cases}
\end{align}



\end{document}